\documentclass[3p,11pt]{elsarticle}

\usepackage{hyperref}

\usepackage{adjustbox}
\usepackage{multirow}
\usepackage{mwe}
\usepackage{xcolor}
\usepackage{wrapfig}
\usepackage{lscape}
\usepackage{rotating}
\usepackage{array, makecell}
\usepackage{floatrow}
\usepackage{amsmath,newtxtext,newtxmath}

\usepackage{graphicx}
\graphicspath{ {images/} }
\usepackage{amsmath}
\usepackage{tabularx, booktabs}
\usepackage{siunitx}
\usepackage{breqn}
\usepackage{url}
\usepackage{subcaption}
\usepackage{bm}

\usepackage{setspace}
\journal{Energies}

\newcommand{\fref}[1]{Fig.~\ref{#1}}
\newcommand{\sref}[1]{Section~\ref{#1}}
\newcommand{\srefs}[2]{Sections~\ref{#1} and \ref{#2}}
\newcommand{\tref}[1]{Table~\ref{#1}}
\newcommand{\eref}[1]{(\ref{#1})}
\newcommand{\sensmat}[0]{SM}

\begin{document}

\begin{frontmatter}

\title{A Sensitivity Matrix Approach Using Two-Stage Optimization for Voltage Regulation of LV Networks with High PV Penetration}



\author[mymainaddress]{A.S. Jameel Hassan\corref{correspondingauthor}\fnref{contrib}}
\ead{jameel.hassan.2014@eng.pdn.ac.lk}

\author[mymainaddress]{Umar Marikkar\fnref{contrib}}
\ead{umar.m@eng.pdn.ac.lk}

\cortext[correspondingauthor]{Corresponding author}
\fntext[contrib]{Equally contributing authors}

\author[mymainaddress]{G.W. Kasun Prabhath}
\ead{gwkprabhath@eng.pdn.ac.lk}

\author[mymainaddress]{Aranee Balachandran}
\ead{aranee.balachandran@eng.pdn.ac.lk}

\author[mymainaddress]{W.G. Chaminda Bandara}
\ead{chaminda.bandara@eng.pdn.ac.lk}

\author[mymainaddress]{Parakrama B. Ekanayake}
\ead{mpb.ekanayake@ee.pdn.ac.lk}

\author[mymainaddress]{Roshan I. Godaliyadda}
\ead{roshangodd@ee.pdn.ac.lk}

\author[mymainaddress,mysecondaryaddress]{Janaka B. Ekanayake}
\ead{ekanayakej@eng.pdn.ac.lk}

\address[mymainaddress]{Department of Electrical and Electronic Engineering, University of Peradeniya, Sri Lanka}
\address[mysecondaryaddress]{School of Engineering, Cardiff University, UK}

\begin{abstract}
The occurrence of voltage violations are a major deterrent for absorbing more roof-top solar power to smart Low Voltage Distribution Grids (LVDG). Recent studies have focused on decentralized control methods to solve this problem due to the high computational time in performing load flows in centralized control techniques. To address this issue a novel sensitivity matrix is developed to estimate voltages of the network by replacing load flow simulations. In this paper, a Centralized Active, Reactive Power Management System (CARPMS) is proposed to optimally utilize the reactive power capability of smart photo-voltaic inverters with minimal active power curtailment to mitigate the voltage violation problem. The developed sensitivity matrix is able to reduce the time consumed by 48\% compared to load flow simulations, enabling near real-time control optimization. Given the large solution space of power systems, a novel two-stage optimization is proposed, where the solution space is narrowed down by a Feasible Region Search (FRS) step, followed by Particle Swarm Optimization (PSO). The performance of the proposed methodology is analyzed in comparison to the load flow method to demonstrate the accuracy and the capability of the optimization algorithm to mitigate voltage violations in near real-time. The deviation of mean voltages of the proposed methodology from load flow method was; $6.5 \times 10^{-3}$ p.u for reactive power control using Q-injection, $1.02 \times 10^{-2}$ p.u for reactive power control using Q-absorption, and 0 p.u for active power curtailment case.

\end{abstract}


\end{frontmatter}


\section{Introduction}

Over the years, the integration of renewable Distributed Energy Resources (DER) to Low Voltage Distribution Grids (LVDG) has gained high prominence due to technological advancements, increased demand in sustainable energy resources and the advent of de-carbonisation programs by many countries \cite{ref25}\nocite{ref38}-\cite{energy_2}. In light of the increase in DER, photovoltaic (PV) generation systems are shown to be the most effective DER prospective for LVDGs \cite{ref26}. However, since the conventional LVDG was designed based on the assumption that power flow would be from the primary substation to the loads \cite{ref1}, high PV penetration gives rise to unforeseen problems \cite{akeyo2021study}. The high penetration of rooftop-PV in LVDGs can result in reverse power flows \cite{Ma_Desenbrock2019} and an increase in the neutral current, leading to distribution and transformer losses due to overheating of the conductor \cite{ref1},\cite{ref2}\nocite{pvrephasing,Ma2019}-\cite{Yaghoobi2018}. A major problem of reverse power flow is the occurrence of upper limit voltage violations, where the busbar voltage at specific points of LVDGs is greater than the specified limit \cite{ref3_1}. Further, studies reveal that voltage violations can occur at a penetration level as low as 2.5\% due to integration of rooftop PV panels at prosumers will \cite{aziz2017pv}. If such voltage violations occur over sustained periods of time it will cause severe damages to loads connected to LVDGs. These detrimental effects of voltage violations compel the utility providers to limit the usable PV capacity for LVDGs. Therefore, there exists a crucial need for an effective solution to encourage the future integration of PV to LVDGs by attempting to mitigate the quality-of-supply ramifications. How to mitigate the voltage violations in LVDGs is a long-standing question to which much time and study has been devoted.\par

Multiple methods have been proposed in the literature to overcome this problem of voltage violations in LVDGs. Feeder enhancement is one such method based on changing the feeder cable with a larger cable or changing the characteristics of the feeder, such as changing the values of multi-grounded resistances \cite{shahnia2011voltage}. While this improves the voltage limits while decreasing neutral current, the approach is highly expensive. Moreover, given the future consumption and PV penetration possibilities, this is not the most economical solution. A more viable solution is the use of On Load Tap Changing (OLTC) transformers to change the tap positions to control the voltage levels \cite{ref3}\nocite{ref9, ref13}-\cite{payne2021dynamics}. But, since frequent tap changes can increase the stress on the transformer, hence reducing its lifespan; a novel optimization algorithm is proposed for resource sharing in \cite{xie2019coordinated} to reduce the tap changing operations. However, the drawback of slow response speed in OLTC switching persists. In order to remedy this issue, fast response devices such as Battery Energy Storage Systems (BESS) and STATCOMs can be installed \cite{ref28}\nocite{arshad2018monte,Zhao2020,MaYiju2019}-\cite{alzahrani2020minimization}. A piece-wise droop control using BESS for rapid changes in voltage profiles is presented in \cite{mak2018hierarchical}. More recently, a reinforcement learning based management technique of BESSs is introduced in \cite{al2020reinforcement}. \par

A more promising control method is the use of Active Power Curtailment (APC) during high PV penetration \cite{ref30}\nocite{ref31,ref14}-\cite{Howlader2020}. Due to the higher impact on voltage profiles by nodes at the farther end of the feeder, most APC operations are performed on distant customers. Since this is not equitable, a fair prosumer based APC approach is proposed in \cite{latif2016quantification}. A novel approach incorporating the Self-Consumption Ratio (SCR) of the customer to determine the allowable PV injection is developed in \cite{Nousdilis2018}. Despite the effectiveness, the spilling of solar power is not an economically attractive solution. Moreover, it is a waste and also detrimental to the whole purpose of renewable energy usage which is to improve the energy mix such that the renewables receive a larger chunk. A more comprehensive solution to this problem is to utilize the capability of the PV inverters to the fullest to supply reactive power in order to mitigate voltage violations. Whilst this is a cost effective method requiring no additional installations, mitigating voltage violations in the 3-phase unbalanced system using only Reactive Power Control (RPC) is a challenging problem \cite{ref32,RPI-APCdroop}. Due to the large R/X ratios of distribution networks, the effect of reactive power control is limited. Therefore to completely remove the violations in the upper limit, the APC is required.\par

Recent studies have vastly explored the APC and RPC mechanisms to minimize voltage violations. These studies can be categorised into two; local/decentralized control and centralized control methods. Control actions of decentralized control methods rely completely on local measurements \cite{ref12}\nocite{ref19, Wang2018, Gandhi2020}-\cite{Tina2019}. A combined approach of RPC and APC as a droop control mechanism to mitigate the voltage violations is proposed in \cite{RPI-APCdroop}. A Volt-Var control (VVC) using two methods to determine the reactive power equation slope is given in \cite{ref20}. It presents a method with the robust minimization of absolute voltage deviation, and a close formed solution inspired from chance constraints. In \cite{Zhang2021} a rule based decentralized RPC is performed taking into account the most sensitive nodes in the network. An optimization technique is developed in \cite{ref21} to coordinate fast dispatch of PV inverters with OLTCs in a decentralized manner due to computational burden in centralized systems. Meanwhile, a two-level control algorithm incorporating OLTCs and BESSs with decentralized RPC is proposed in \cite{Emarati2021}. Multiple works in the literature have also developed control mechanisms based on droop control \cite{ref22}-\cite{singh2020multistage}. Nevertheless, the lack of information about the entire network status in decentralized control prevents the optimal use of reactive power capacity in controlling the voltage violations.\par

\begin{figure}[htb]
    \centering
    \includegraphics[width=.7\textwidth]{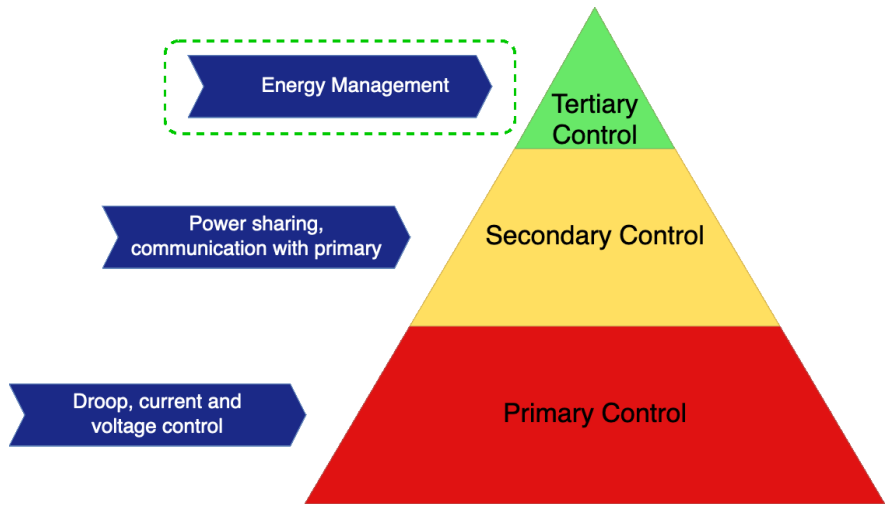}
    \caption{Hierarchical control architecture.}
    \label{fig:control_hierarchy}
\end{figure}

However, provided that sufficient information about the network can be retained, centralized control is more efficient compared to decentralized control \cite{ilea2020voltage}. Such network state observability is achieved by means of solar predictions \cite{MAR_ours} and state estimation \cite{State_est} enabling a control at the tertiary level of the control architecture as shown in Fig.\ref{fig:control_hierarchy}. To overcome the lack of information of the network, a global solution is attained by the centralized control method which determines the power injections/absorptions/curtailment by means of an Optimal Power Flow (OPF) problem \cite{ref2_20}. In \cite{ref8}, a comprehensive PV control strategy is proposed to improve the operational performance of significantly unbalanced 3 phase 4 wire LVDG with high residential PV penetration, by converting a multi-objective OPF problem into a single objective OPF problem. A control algorithm is introduced for maintaining the average customer voltage profile obtained before introducing the PV into the circuit using the control of automatic devices, such as voltage regulation and switched capacitor banks along with PV inverter reactive power \cite{ref18}. Here, the PV inverter control settings are determined by the circuit loading, time of day and PV location in the network. A combination of centralized and decentralized control strategies utilising OLTCs and Capacitor Banks (CB) is also proposed in \cite{Ma2021,Wang2019}. It further analyses the impact on the substation end and the effect of unbalance in phases in PV integration. \par

However, these methods suffer from high computation time due to varying reasons such as the need to solve load flows within the optimization algorithm and the integration of VAR compensation equipment. Most of the referred centralized methods related to power systems control use load flow analysis to calculate voltage variation \cite{ref4_4} \nocite{Mahmoud2020}-\cite{ref4_1}. Since these methods achieve accurate results at the expense of time, a voltage and PV-power sensitivity approach is used to calculate the voltage variations \cite{SensMat_invNR1}\nocite{SensMat_invNR2}\nocite{zhuSensMat,SensMat_surfaceplot}-\cite{SensMat_MV}. The different sensitivity matrices used in the literature are discussed in Table 1.\par

\begin{table}[htb]
    \centering
    \caption{\fontsize{10}{10}\selectfont Sensitivity matrices existing in the literature}
    
    \begin{tabular}{m{0.38\textwidth} >{\centering\arraybackslash}m{0.15\textwidth} m{0.4\textwidth}}

    \toprule
    \textbf{How the sensitivity matrix was developed} & \textbf{References} & \textbf{Disadvantages of the method}\\
    \midrule
    Inverse from Jacobian of Newton Rhapston power flow equations & \cite{SensMat_invNR1,SensMat_invNR2} &   Repetitive computation of the inverse of the Jacobian which is computationally expensive with the increase in matrix size.\\
    \midrule
    Surface fitting technique and using simulations of multiple load flow analysis &  \cite{zhuSensMat,SensMat_surfaceplot} & An extensive simulation needs to be run in case of a change in the network parameters to be able to develop a new sensitivity matrix that will fit the network.\\
    \midrule
    Using the topological structure of the network &    \cite{SensMat_MV} & The derivation is done for a MV distribution line assuming constant voltage for the slack bus. However, the secondary voltage of the LV network will fluctuate, which needs to be accounted for.\\
    \bottomrule
    
    \end{tabular}
    \label{tab:sens mat}
\end{table}

In order to find the optimum solution to the centralized control method in mitigating voltage violations, many optimization techniques have been experimented. Among these SQP\cite{ref8}, Non-Linear Programming (NLP)\cite{ref2_19, ge2020decoupling}, Evolutionary Algorithm \cite{alboaouh2018voltage}, Langrangian multipliers \cite{de2020coordinated}, Multi-Objective Evolutionary Algorithm (MOEA) \cite{Ferreira2020} and Particle Swarm Optimization (PSO) \cite{ref4_4},\cite{PSO} algorithms were widely used. In order to act as a viable near real-time system, the accuracy and the computational time of the algorithm plays a key role. Given the vast solution space of LVDG networks, i.e: high complexity of the network due to the number of PV connections in the power system, the computational time for convergence grows dramatically. Therefore, optimization techniques need to be tailored for LVDG power systems such that the computational time is minimal whilst maintaining robustness in terms of convergence to the optimal solution. \par

In this paper, we propose a Centralized Active Reactive Power Management System (CARPMS) which uses the combination of both RPC and APC to mitigate the voltage violations in LVDGs in the tertiary control level. A sensitivity matrix derivation for voltage with respect to the PV power changes and a modified two-stage optimization process with a Feasible Region Search (FRS) and PSO, to find the optimal power settings were developed. The incorporation of the sensitivity matrix vastly reduced the computational time as compared to traditional load flow based optimization in centralized control. In addition, the FRS step in the two-stage optimization process was able to greatly reduce the search space for the solution by narrowing the solution towards the optimum and decreasing the time, thereby enabling a real-time application of the proposed solution. The PSO algorithm was used as the second step to drive the solution to its best solution to prevent frequent violations in the network. \par 

The proposed CARPMS was simulated on a network belonging to an existing housing complex named 'Lotus Grove', located in Colombo, Sri Lanka. The case study network was chosen from the same region of the authors whilst being similar to the IEEE European low voltage test feeder \cite{pes_test_feeder, boglou2020fuzzy} in network size and topology. Specifically, the following contributions are made in this paper:
\begin{itemize}
    \item A novel PV-power to voltage Sensitivity Matrix (\sensmat) for LVDGs is developed using line parameters accounting for the voltage variations in the secondary side of the transformer.
    \item A Centralized Active Reactive Power Management System (CARPMS) using this \sensmat\ for voltage violations in LVDGs is proposed.
    \item A modified two-stage optimization process with the Feasible Region Search (FRS) as the first step is developed to reduce the search space and decrease the computational time. The PSO is used as the second step to find the optimal solution using the FRS solution. 
\end{itemize}

\section{Methodology}

\subsection{Centralized Active Reactive Power Management System (CARPMS)}

\begin{figure}[bth]
\begin{center}
\includegraphics[width=0.8\textwidth]{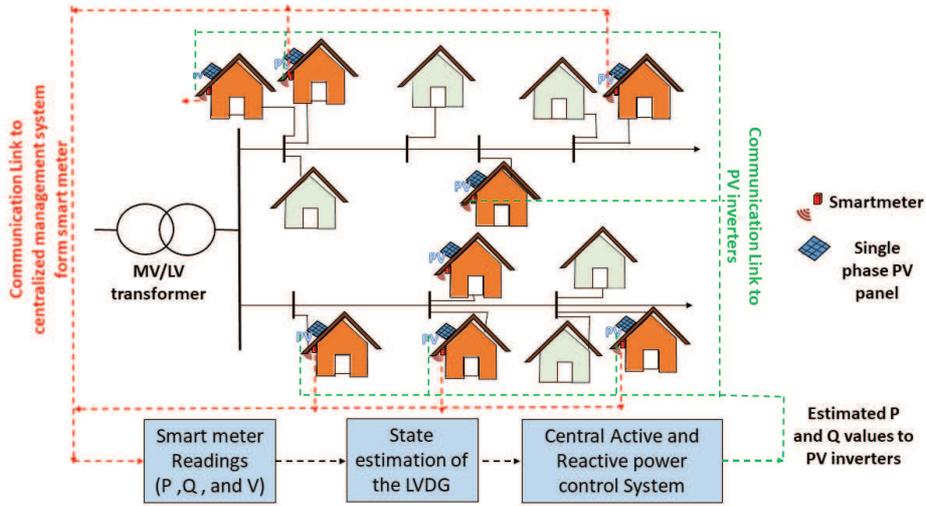}
\caption{Schematic overview of the CARPMS information flow}
\label{fig:smartgrid}
\end{center}
\end{figure}

In this section, the proposed Centralized Active Reactive Power Management System (CARPMS) which eliminates the voltage limit violations at each node is described. Fig.\ref{fig:smartgrid} shows the operating mechanism of the CARPMS. The CARPMS is equipped with smart meters at each PV panel in the network, ensuring the access to active, reactive power and voltage readings at each PV panel node. Due to delays incurred in communication and algorithm processing time \cite{ref2_14} real-time data will not reach the CARPMS. Therefore it acts as a real time management system with control actions relying on estimations of the network states predicted using historical data \cite{State_est}. The proposed algorithm described in \sref{sec:optim} is then used by the CARPMS to detect and correct any voltage violations in the nodes. \par

\begin{figure}[H]
\centering
\includegraphics[width=.75\textwidth]{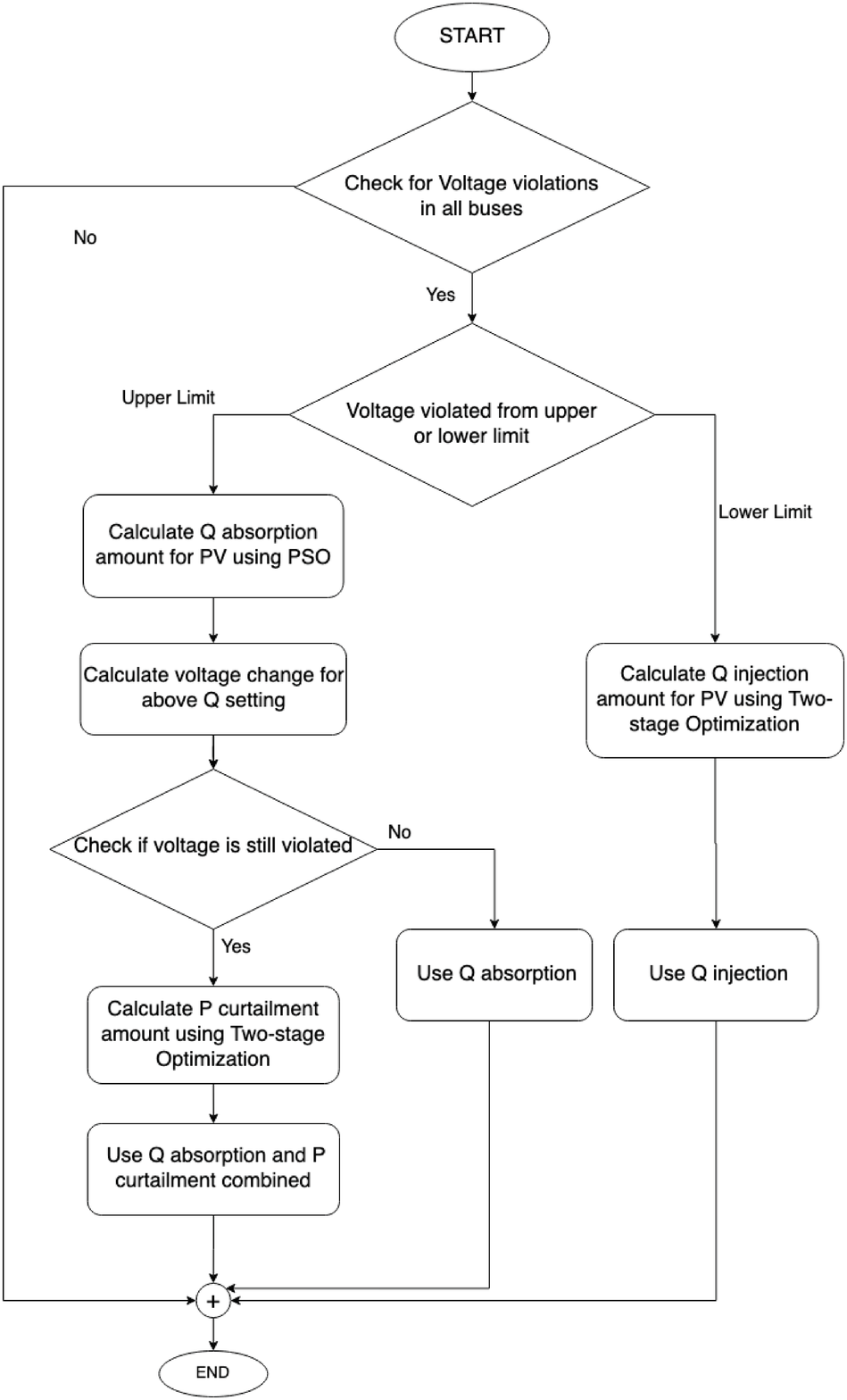}
\caption{Proposed algorithm steps}
\label{fig:algo flowchart}
\end{figure}

The proposed algorithm will encounter voltage violations of two-fold: upper limit and lower limit violations. Due to the low X/R ratio, the violations cannot be entirely removed by RPC alone. In this case, the algorithm utilises an optimized combination of RPC and APC. A detailed flow of the algorithm steps is highlighted in Fig.\ref{fig:algo flowchart}. The derivation of the \sensmat\  used is given in Sections \ref{subsec:VoltageSens Distrb line}-\ref{subsec:Combined sens model} and the two-stage optimization in the control algorithm is described in \srefs{sec:Problem formulation}{sec:optim}.

\subsection{Voltage sensitivity derivation for distribution line}
\label{subsec:VoltageSens Distrb line}

The \sensmat\  is derived for a network without sparse line connections. This assumption is made for the ease of proof which can be easily extended for a network with sparse if necessary. \par

\begin{figure}[htb]
\begin{center}
\includegraphics[width=8cm]{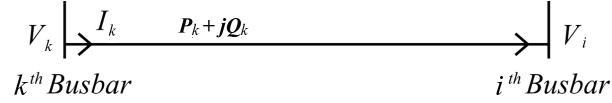}
\caption{A distribution line}
\label{fig:distribution line}
\end{center}
\end{figure}

Consider a phase of a distribution line shown in Figure \ref{fig:distribution line}. Due to the negligible effect of the longitudinal component, by neglecting the power losses, the voltage drop between the $k^{th}$ and $i^{th}$ busbar is given by,
\vspace{-0.5cm}

\begin{equation}
         |V_{k} - V_{i}| \cong \frac{P_{k}R_{k}+Q_{k}X_{k}}{|V_{k}^*|} 
\label{eq:voltage approx}
\end{equation}

where, \bm{$V_{k}$}, \bm{$V_{i}$} are the complex voltages at $k^{th}$ and $i^{th}$ busbars respectively, $R_{k}$ is the resistance of the line, $X_{k}$  is the reactance of the line, $P_{k}$ is the active power flow through the line and $Q_{k}$ is the reactive power flow through the line.

The equation above expresses the difference in the magnitude of the voltage between two adjacent busbars. This equation was extended to calculate the voltages of all the busbars in the network. To generalise, a radial LVDG network with $(N+1)$ number of busbars was considered. Considering the power flow from the LV transformer in the network as positive power flow and using \eref{eq:voltage approx}, the voltage drop up to $r^{th}$ busbar from the transformer end is given by,
\vspace{-0.5cm}

\begin{equation}
         |V_{0}-V_{r}| = \sum_{h=0}^{r-1} \frac{P_{h}R_{h}+Q_{h}X_{h}}{|V_{h}^*|} 
\label{eq:r_th bus drop}
\end{equation}

where, $V_{0}$ is the secondary voltage of the transformer for one of the three phases $a$, $b$ or $c$, which is also the zero\textsuperscript{th} busbar of the network.\par

The power flow of the transmission line is a collective function of domestic loads, PV generations and power transmission losses. However, the power transmission losses are negligible compared to other variables. Thus, the power transmitted through the transmission line was derived as follows, which can be substituted in \eref{eq:r_th bus drop} yielding,
\vspace{-1cm}

\begin{align}
    P_{h}+jQ_{h} &= \sum_{m=h+1}^{N} \left((P_{L_m}-P_{PV_m}) +j(Q_{L_m}-Q_{PV_m})\right)  \label{eq:busPQ}\\
    |V_{0}-V_{r}| = \sum_{h=0}^{r-1} &\frac{\sum_{m=h+1}^{N}\left( (P_{L_m}-P_{PV_m})R_{h}+(Q_{L_m}-Q_{PV_m})X_{h}\right)}{|V_{h}^*|}  \label{eq:r_th bus expanded}
\end{align}

\begin{figure}[htb]
\begin{center}
\includegraphics[width=0.6\textwidth]{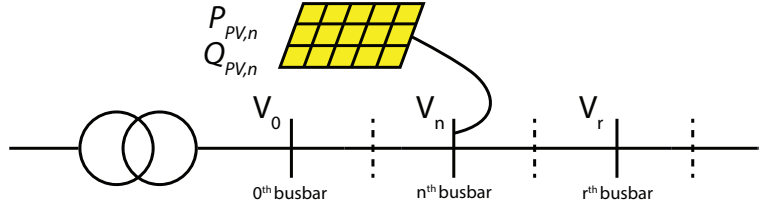}
\caption{A schematic LVDG network}
\label{fig:PVsetting}
\end{center}
\end{figure}

The network parameters $X_{h}$ and $R_{h}$ given in \eref{eq:r_th bus expanded} are constant, unique and attainable for every LVDG network. Whilst the load power and PV power generation parameters are not easily obtainable in real-time, estimation of these parameters are possible \cite{ref2_22}-\nocite{ref2_23}\nocite{ref2_24}\nocite{ref2_25}\cite{ref2_28}. The derivation of voltage of $r^{th}$ busbar with respect to the reactive power of the PV system in the $n^{th}$ busbar was derived as,
\vspace{-1cm}

\begin{align}
    \frac{\partial V_{r}}{\partial Q_{PV_n}} &= \frac{\partial V_{0}}{\partial Q_{PV_n}}+\sum_{h=0}^{n-1} \frac{ X_{h}}{|V_{h}^*|} \hspace{100pt} for (r\geq n)  \label{equ7}\\  
    \frac{\partial V_{r}}{\partial Q_{PV_n}} &= \frac{\partial V_{0}}{\partial Q_{PV_n}}+\sum_{h=0}^{r-1} \frac{ X_{h}}{|V_{h}^*|}     \hspace{100pt} for (r<n) \label{equ9}
\end{align}

where, $\frac{\partial V_{r}}{\partial Q_{PV_n}}$ is voltage sensitivity of the $r^{th}$ busbar with respect to the reactive power variation of the PV panel at the $n^{th}$ busbar and $V_{0}$ is the voltage of the busbar connected at the secondary side of the transformer. A schematic LVDG network showing the busbar notations is shown in \fref{fig:PVsetting}.

Similarly, the voltage sensitivity of busbars with respect to the active power of the PV system in the $n^{th}$ busbar can be derived. 

\subsection{Voltage sensitivity derivation at the transformer end}
\label{subsec:VoltageSens transf end}

\begin{figure}[htb]
\begin{center}
\includegraphics[width=0.6\textwidth]{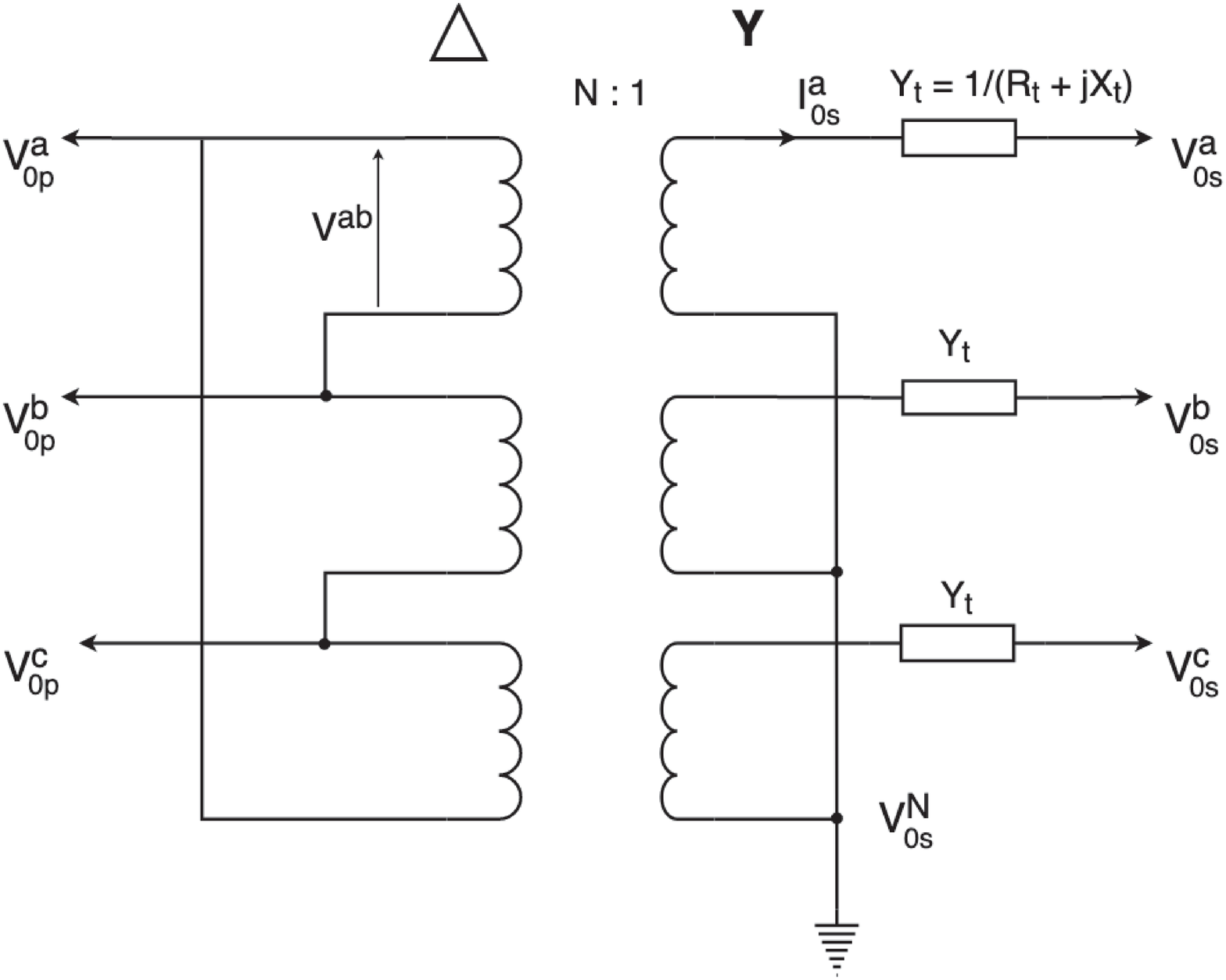}
\caption{Equivalent circuit of the Delta-Wye transformer}
\label{fig2}
\end{center}
\end{figure}

In order to calculate $\frac{\partial V_{0}}{\partial Q_{PV_n}}$ and similarly and $\frac{\partial V_{0}}{\partial P_{PV_n}}$ the LV transformer of the residential network was modeled as shown in Figure \ref{fig2}.

Considering the secondary side of the transformer, the expression for current and power flow in the secondary side were derived in terms of voltages and impedances using the transformer model matrix in \cite{JBE_SG}. The power flow in the secondary side of the transformer was obtained as,
\vspace{-0.7cm}

\begin{equation}
P_{s}^a-jQ_{s}^a = (V_{0_p}^a \cdot Y_1-V_{0_p}^b \cdot Y_1+V_{0_s}^N \cdot Y_2)*|V_{0_s}^{a}|\angle{-\delta_k}  -|V_{0_s}^a|^2 \cdot Y_2  
\label{equ13}
\end{equation}
where, $V_{0_p}^a$, $V_{0_p}^b$ and $V_{0_p}^c$ are the primary side voltages of the LV transformer, $V_{0_s}^a$, $V_{0_s}^b$ and $V_{0_s}^c$ (generically denoted by $V_{0}$ in previous section) are the secondary side voltages of the LV transformer and $I_{0_s}^a$, $I_{0_s}^b$ and $I_{0_s}^c$ are the secondary side currents of the LV transformer of $a$, $b$ and $c$ phases respectively. $Y_1$ and $Y_2$ are the primary and secondary side impedance of the LV transformer and $N$ is the secondary to primary transformer turns ratio. $ Y_1=\frac{Y_t}{N}$ and  $Y_2=Y_t $, $V_{0_s}^{a}=|V_{0_s}^{a}|\angle \delta _a $. \par

In order to calculate the voltage sensitivity of the transformer end with respect to reactive power changes, the derivative of the imaginary component of \eref{equ13} with respect to $V_{0_s}\textsuperscript{a}$ was obtained as,
\vspace{-0.7cm}

\begin{equation}
\frac{\partial Q_{s}^a}{\partial |V_{0_s}^a|} = 2|V_{0_s}^a| \cdot img(Y_2) -img((V_{0_p}^a \cdot Y_1 - V_{0_p}^b \cdot Y_1 + V_{0_s}^N \cdot Y_2)\angle{-\delta_a}) 
\label{eq:V0sensQ}
\end{equation}

Since the active and reactive power of loads can be assumed to be constant during a control sequence operation, using \eref{eq:busPQ} the variation of reactive power for a given phase with respect to the transformer end voltage is only the variation of PV power in that phase.

Then by obtaining the reciprocals, the variation of transformer end voltage with respect to PV reactive power connected to the given phase was computed using equation (\ref{eq:V0sensQ}).
\vspace{-0.5cm}

\begin{equation}
    \frac{\partial |V_{0_s}^{a}|}{\partial Q^{a}_{{PV}}} = \frac{\partial |V_{0_s}^a|}{\partial Q_{s}^a} 
    \label{eq:Vo/Qpv}
\end{equation}

Similarly, using the real part of \eref{equ13}, the variation of transformer end voltage with respect to PV active power can be obtained.

\subsection{Combined Sensitiviy Matrix model}
\label{subsec:Combined sens model}

The combined sensitivity model was derived based on the results from \srefs{subsec:VoltageSens Distrb line}{subsec:VoltageSens transf end}. Considering the number of PV panels in the system as $M$ and the number of busbars in the system as $N$, using Equations (\ref{equ7}), (\ref{equ9}), (\ref{eq:Vo/Qpv}) and their analogous equations for active power, the combined \sensmat\ model of the network with respect to the power generation of PV systems was derived as,  
\vspace{-0.7cm}

\begin{equation}\label{eq:Combined Sens model}
\begin{split}
    \begin{bmatrix} \Delta V  \end{bmatrix}_{N \times 1}
    &=
    \begin{bmatrix}
        \frac{\partial V}{\partial Q_{PV}} & \frac{\partial V}{\partial P_{PV}} 
    \end{bmatrix}_{N \times 2M}
    \begin{bmatrix}  
        \Delta  Q_{PV} \\ \Delta  P_{PV}
    \end{bmatrix}_{2M \times 1}\\
    &= \begin{bmatrix} 
        \sum_{h}\frac{X_{h}}{|V_{h}^{*}|} & \sum_{h}\frac{R_{h}}{|V_{h}^{*}|}
    \end{bmatrix}_{N \times 2M}
    \begin{bmatrix}
        \Delta Q_{PV}\\
        \Delta P_{PV}
    \end{bmatrix}_{2M \times 1}
    + \begin{bmatrix} \Delta V_{0} \end{bmatrix}_{N \times 1}
\end{split}
\end{equation}

where,$\begin{bmatrix} \Delta V_0  \end{bmatrix}_{N \times 1}$ is the voltage change at the LV transformer end and $\begin{bmatrix} \Delta V  \end{bmatrix}_{N \times 1}$ is the combined voltage variation at each busbar due to the PV power variations (at each iteration). The system has been linearised assuming that the variation in the PV system power within the control sequence algorithm to be considerably small.

\section{Problem formulation}
\label{sec:Problem formulation}
The aim of this work is to determine an optimum setting to prevent voltage violations in LV networks. An RPC mechanism followed by APC is carried out if RPC alone is unable to rectify the voltage violations. Thus, two optimization functions defined in \srefs{subsec:RPC optimization}{subsec:APC optimization} were proposed to converge on the optimal operating point. The objective functions of the optimization algorithms intend to minimize the active and reactive power settings whilst satisfying the voltage limit constraints and the inverter constraints. The voltage limits pertain to the lower and upper limits of the acceptable voltages in LVDG networks, whereas the inverter constraints depend on the power ratings of the inverters. \par

\begin{figure}[htb]
\begin{center}
\includegraphics[width=0.9\textwidth]{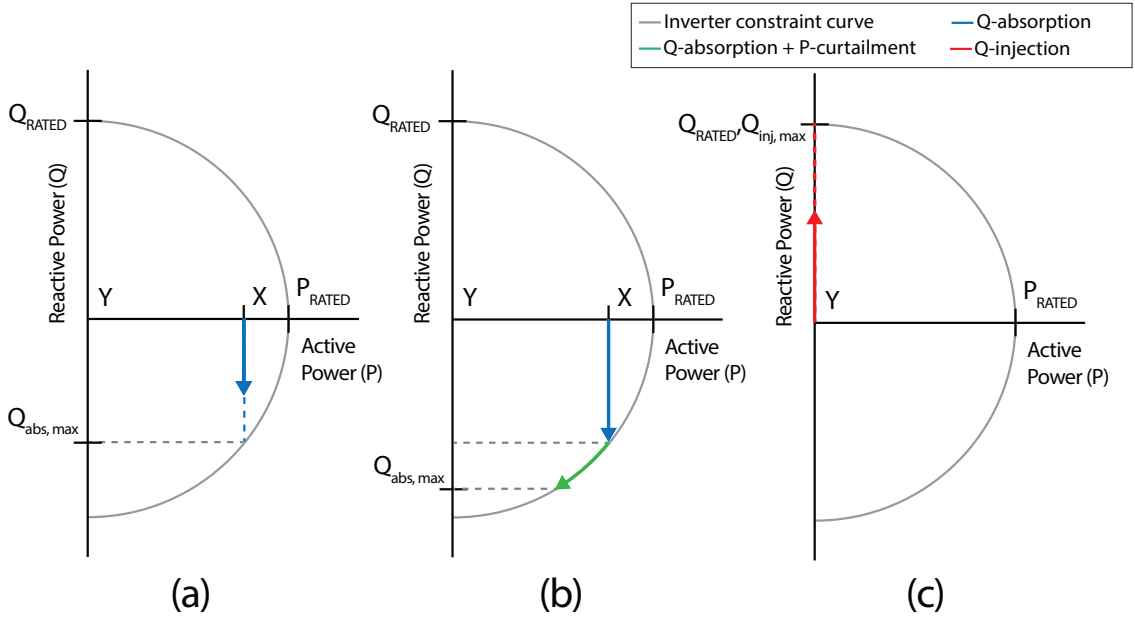}
\caption{PV inverter status change for control using (a) Q-absorption, (b) Q-absorption and P-curtailment and (c) Q-injection.}
\label{fig:inverter_cons}
\end{center}
\end{figure}

The state of the PV inverter being varied by the algorithm depending on the 3 control methods Q-absorption, P-curtailment and Q-injection is depicted in \fref{fig:inverter_cons}. The "X" mark shows an instance of an initial state of the inverter during the day. During RPC Q-absorption, the state moves vertically downwards upto a given optimum point. If it reaches the inverter constraint/capability curve, it implies that Q-absorption cannot be performed under the given conditions. Then, P-curtailment is performed during which the state of the inverter moves along the capability curve, reducing the amount of active power injected to the network. It can also be noted that the maximum allowable value of Q-absorption: $Q_{abs,max}$, varies depending on the active power state of the inverter. The inverter state during night-time is marked "Y". Here, RPC (Q injection) is carried out and the inverter state moves upwards along the Q-axis (injecting reactive power to the network) till it reaches an optimal point, or its full Q injection capacity: $Q_{inj,max}$. \par

\subsection{Optimization of Reactive Power Control}
\label{subsec:RPC optimization}
The objective function of the optimization of RPC was expressed as a function of the total deviation of busbar voltages from 1 p.u and the neutral voltage as given by,
\vspace{-0.7cm}

\begin{equation}\label{eq:objfuncQ}
    J_{RPC} =\text{min} \sum_{i=1}^{n} \left(c_{d}*V_{d,i} + c_{neut}*V_{neut,i} \right)
\end{equation}
where $V_{d,i}$ is the total deviation of voltages of busbar $i$ from 1 p.u, $V_{neut}$ is the neutral voltage of busbar $i$, and $c_{d}$ and $c_{neut}$ are scaling constants.\par

\noindent \textbf{Subjected to the constraints:}
\begin{enumerate}
    \item The voltage of the busbar should be within the specified upper and lower limits given by,
    \vspace{-0.7cm}

    \begin{equation}
    V_{\text{lower limit}} \leq V_{\text{buses}} + \Delta V  \leq V_{\text{upper limit}}
    \label{eq:voltconstrQ}
    \end{equation}

    where, $V_{\text{lower limit}}$ and $V_{\text{upper limit}}$ are the accepted lower(0.95 p.u) and upper limit(1.05 p.u) voltages in LVDG systems respectively, $V_{\text{buses}}$ is the calculated voltage of the busbars using optimization variables and $\Delta V$ is the estimated voltage change due to changes in $P$ and $Q$ of the PV systems.
    
    \item The inverter constraints given below should be satisfied,
    \vspace{-0.7cm}

    \begin{equation}
    S_{PV_i}^2 \geq  P_{PV_i}^2+Q_{PV_i}^2
    \label{eq:inverter constrQ}
    \end{equation}
\end{enumerate}

The variables of the optimization problem were the reactive power setting at each busbar with a PV panel in the network which can be expressed as, \emph{$\text{Optim. Variable (OV) } = \left[ Q_{pv_1},Q_{pv_2},Q_{pv_3},Q_{pv_4},........,Q_{pv_m} \right]$}.

In order to formulate this optimization problem to minimize \eref{eq:objfuncQ} whilst satisfying the above constraints, penalties were introduced to ensure that the optimal solution satisfies the constraint of voltage violations to the best case possible by penalising the cost function when constraints are violated. Hence, the optimization problem is reformulated as a minimisation of the penalised objective function $J_{1}$ given by \eref{eq:objfuncQPEN}.
\vspace{-0.7cm}

\begin{equation}\label{eq:objfuncQPEN}
J_{1} = \text{min}\ \left(c_{vial}*e^{N_{vial}}*J_{RPC}\right)
\end{equation}
where $N_{vial}$ is the sum of the number of violations in each phase and $c_{vial}$ is a scaling constant.

\subsection{Optimization of Active Power Curtailment}
\label{subsec:APC optimization}
The objective function of the optimization of the APC included the amount of active power curtailed and was expressed as,
\vspace{-0.7cm}

\begin{equation}
J_{APC} =\text{min} \sum_{i=1}^{n} \left( \Delta P_{PV_i} + c_{d}*V_{d,i} + c_{neut}*V_{neut,i} \right)
\label{eq:objfuncP}
\end{equation}
where $\Delta P_{PV_{i}}$ is the amount of curtailed active power, $V_{d,i}$ is the total deviation of voltage of busbar $i$  from 1 p.u, $V_{neut}$ is the neutral voltage of busbar $i$, and $c_{d}$ and $c_{neut}$ are scaling constants.

\noindent \textbf{Subjected to the constraints:}
\begin{enumerate}
    \item The voltage of the buses should be within the specified upper and lower limits as in \eref{eq:voltconstrQ}.
    \item The inverter constraint given below should be satisfied,
    \vspace{-0.7cm}

    \begin{equation}
    S_{PV_i}^2 =  P_{PV_i}^2+Q_{PV_i}^2
    \label{eq:inverter constP}
    \end{equation}
\end{enumerate}

The variables of the optimization problem here were the active power setting at each busbar in the network which can be expressed as, \emph{$\text{Optimization variable (OV)} =[P_{pv_1},P_{pv_2},P_{pv_3},P_{pv_4},........,P_{pv_m}]$}.

Similar to the RPC case, the optimization problem is reformulated by including the same penalties such that the penalty is applied when the constraint is violated. This ensures the optimal solution of the objective function satisfies all constraints.

\section{Two-stage Optimization}
\label{sec:optim}

This section outlines the proposed modified optimization algorithm consisting of two processes in sequence:
\begin{itemize}
    \item Feasible Region Search (FRS)
    \item Particle Swarm Optimization (PSO)
\end{itemize}

The aim of FRS is to drive the elements in the Optimization Variable ($OV$) towards the feasible region, where upper or lower limit voltage violations are non-existent. This is followed by a PSO algorithm, where these variables in the feasible region are then optimized according to a predefined cost function, to find the best possible solution.\par

\subsection{Feasible Region Search}
\label{subsec:FRS}

The number of elements in the $OV$ increases with the number of PV panels connected to a given network, which results in a large search space. At the initial violated conditions, the existence of the $OV$ far away from the feasible region and the high dimensionality of the search space may result in poor performance of a standard PSO algorithm. This is because the first step of the PSO is an initialisation procedure (discussed in \sref{subsec:PSO}) being a random scattering of $OV$s in the neighbourhood of the current $OV$. \par

Through the FRS, we determine a new initial point for the PSO by moving the present $OV$ towards the feasible region. The driving function of the $OV$ is given by,
\vspace{-0.7cm}

\begin{equation}
\label{eq:OViter}
    OV[i,j] = OV[1,j] + \alpha[i] * \Delta OV_{max}[j]
\end{equation}
for all $j=1,2...m\ (\text{total number of elements in } OV[i,:]) $, where $OV[1,j]$, is the initial value of the $j^{th}$ element in the $OV$, driven towards the feasible region by $\alpha[i]$. Here $\alpha[i]$ is the driving parameter which is a monotonically increasing function from 0 to 1. $\Delta OV_{max}[j]$ is the maximum possible change of $OV[:,j]$. This results in $OV[i,j]$, the calculated position of the $j^{th}$ element of the $OV$ at the $i^{th}$ iteration.  \par

In a PV integrated network, the vector $OV[i,:]$ is the power settings of each inverter connected to the network at any given iteration $i$. $\Delta OV_{max}[j]$ is the maximum Q-absorption, Q-injection or P-curtailment capacity of the $j^{th}$ inverter. This also determines the driving direction of elements of $OV$, as the sign of $\Delta OV_{max}[j]$ is dependent on the current nature of voltage violation. In the instance of an upper limit violation, $\Delta OV_{max}[j]$ will be negative, as the P and Q setting of the $j^{th}$ inverter which corresponds to the maximum P-curtailment and Q-absorption capacity is negative. $\Delta OV_{max}[j]$ will be positive for lower limit violations, as the inverter state travels in the direction of positive Q for Q-injection, as discussed in \sref{sec:Problem formulation}. 
\begin{figure}[htb]
    \centering
    \includegraphics[width=10cm]{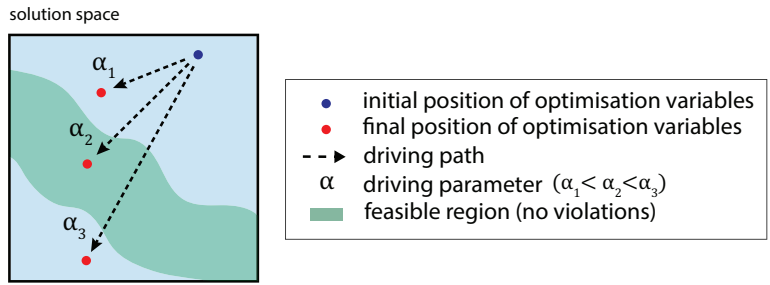}
    \caption{Effect of $\alpha$ on FRS}
    \label{fig:FRS}
\end{figure}

During each iteration, $OV[i,j]$ was computed such that it moves closer to the feasible region. As the sign of $\Delta OV_{max}[j]$ ensures the current $OV[i,:]$ move towards the feasible region, the driving parameter $\alpha$ serves to gradually increase the change in $OV[:,j]$. If this change happens to be very large, there exists a possibility of $OV$ overshooting towards unwarranted solutions, as illustrated in \fref{fig:FRS}. For example, lower limit violations may occur if Q-absorption takes place to its maximum capacity to mitigate an upper limit violation. This problem was overcome by the use of $\alpha$, which gradually increases with each iteration $i$, allowing FRS to terminate as soon as $OV[i,j]$ reaches the feasible region.\par
Upon the termination of FRS, the system is devoid of any upper or lower limit violations. Although it is possible to complete the control process using only FRS, it does not fully optimize the network as it does not consider parameters relative to the cost function given by \eref{eq:objfuncQPEN} and \eref{eq:objfuncP}. Instead it only accounts for the existence of violations in the system. Also, FRS acts as a decoupled control algorithm, where the inverter power settings in the $i^{th}$ iteration are independent of each other, where each element in $OV[i,:]$ is being modified by the same value of $\alpha$.\par

In contrast, although slower than FRS, the PSO algorithm performs as a collective control algorithm. This implies that, the factor $\alpha$ at which Q-absorption, Q-injection or P-curtailment is performed relative to the capacity of the inverter, will be optimized. For instance, in the absence of solar power, PSO will ensure more Q is injected to the network by inverters furthest from the secondary transformer, where the cumulative voltage drop is high. Due to properties of the cost function proposed in \sref{sec:Problem formulation}, PSO will further push the optimal point much more towards the centre of the feasible region. This allows for higher margins of errors in state estimation of the PV integrated network, as a small deviation in the $OV$ will not drift the solution towards unfeasible regions.\par

\subsection{Particle Swarm Optimization}
\label{subsec:PSO}

Particle Swarm Optimization (PSO) is a heuristic algorithm used in problems with high dimensional search domains. It is a nature inspired algorithm which is based on the foraging technique of flocks of birds and schools of fish. There are six steps in standard PSO \citep{PSO},\cite{PSO1}, as shown in Algorithm 1. 

\begin{table}[htb]
    \centering
    \begin{tabular}{p{0.1\textwidth} p{0.83\textwidth}}
    \toprule
    \multicolumn{2}{l}{\textbf{Algorithm 1: Steps of PSO}}    \\
    \toprule
    \textbf{Step 1:} & Initialize swarm of particles (ie: population)\\
    \textbf{Step 2:} & Compute the cost of each particle using fitness function\\
    \textbf{Step 3:} & Record personal best of each particle and global best of entire population\\
    \textbf{Step 4:} & Update velocity of each particle using personal and global best and other parameters\\
    \textbf{Step 5:} & Calculate new position of each particle\\
    \textbf{Step 6:} & Repeat steps 2-5 until the each particle converges to their solution or iteration count is completed, extract global best of the entire population as the optimal solution\\
    \bottomrule
    \end{tabular}
    \label{tab:PSO steps}
\end{table}

Due to the large search space in this problem, the standard PSO is unable to converge to a satisfactory solution. As discussed previously in \sref{subsec:FRS}, FRS is carried out, and the initial population is created by randomly scattering particles in the neighborhood of the $OV$ which is now located in the feasible region.

\subsection{Primary steps of Particle Swarm Optimization}
\label{subsec:PSO steps}
The steps involved in PSO are shown in Algorithm 1. The update equation for the position and velocity of the particles is given by, 
\vspace{-0.7cm}

\begin{equation}
x_i[j+1] = x_i[j] + V_i[ j+1 ] 
\label{eq:PSO pos}
\end{equation}
where $x_i[j]$ denotes the position of the $i^{th}$ particle at the $j^{th}$ iteration, $x_i[j+1]$ and $V_i[j+1]$ are the position and velocity of that particle for the next iteration. The velocity at which the particle travels is expressed by, 
\vspace{-0.7cm}

\begin{equation}
V_i[j+1] = V_i[j] + P_b*(p_i[j] - x_i[j]) + G_b*(g[j] - x_i[j])
\label{eq:PSO vel}
\end{equation}

where, $V_i[j]$ is velocity of the particle at $j^{th}$ iteration, $p_i[j]$ and $g[j]$ denote the current personal and global best of the $j^{th}$ particle, $P_b$ and $G_b$ are the confidence factors for the personal and global best respectively.

The notion of the velocity is to set the direction of search and the extent of exploration by the particle. This depends on where in the search space the current particle exists, the recorded best position of that particle (personal best), and the recorded best position of all particles (global best) since the start of the algorithm. The dependency of the personal or global best on the velocity is governed by confidence factors, expressed by the two variables $P_b$ and $G_b$. These variables were manipulated such that the particles follow their own local optima, for the amount of observed local optima may be high, thus giving a better chance for the global optimum to exist within the observed local optima.
 
\section{Test Network}
\label{sec:Test Network}

\begin{figure}[htb]
    \centering
    \includegraphics[width=0.75\textwidth]{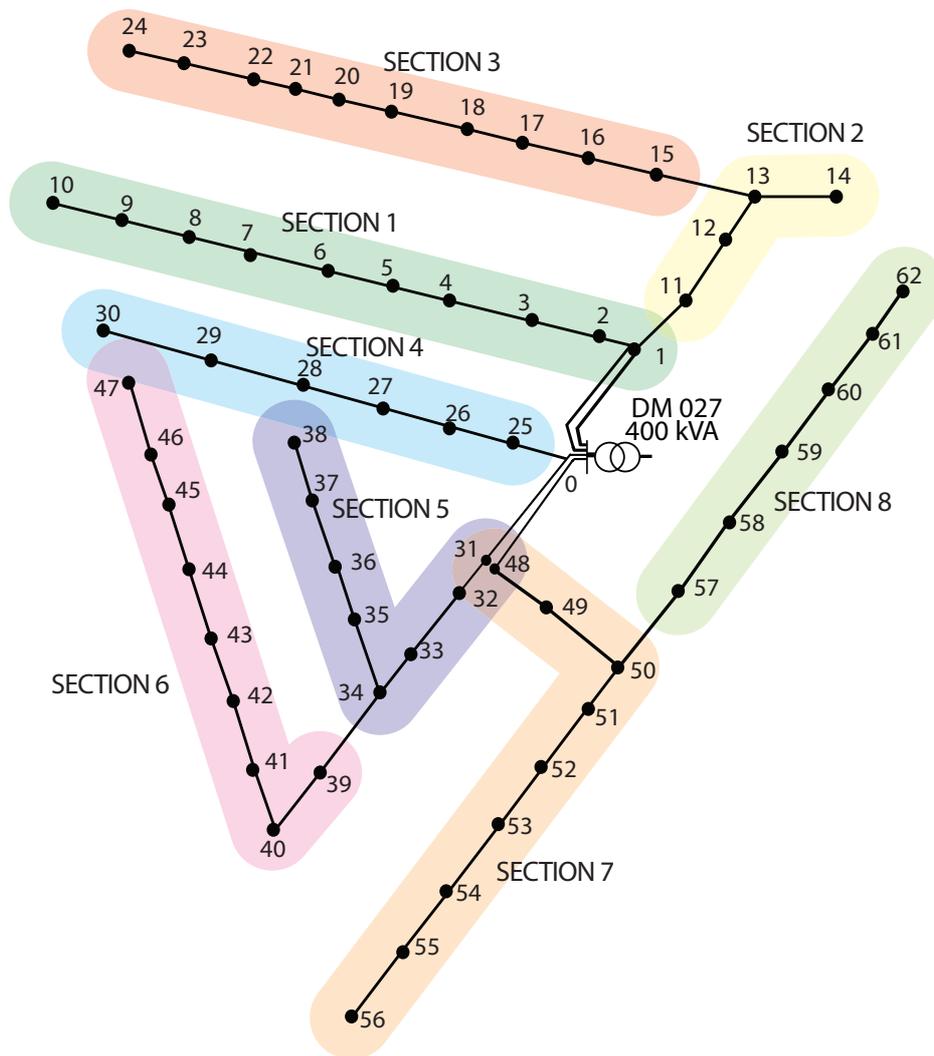}
    \caption{Single line diagram of test LVDG network (Lotus Grove, Sri Lanka) used for simulations}
    \label{fig:network}
\end{figure}

\begin{figure}[!ht]
    \centering
    \includegraphics[width=\textwidth]{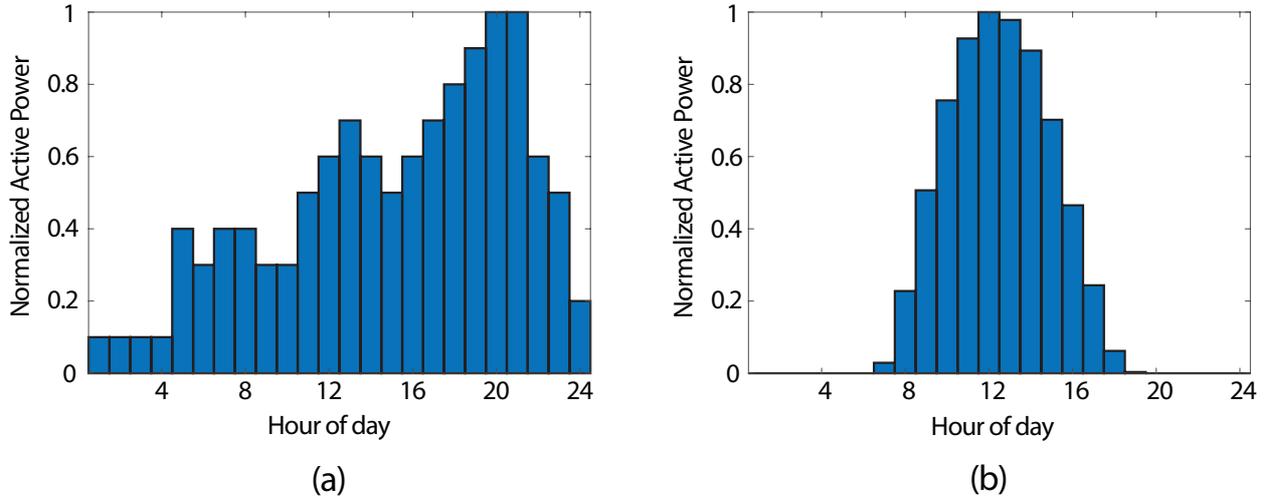}
    \caption{Daily (a)Load profile (b)PV profile}
    \label{fig:dailyprofiles}
\end{figure}

The network belonging to an existing housing complex 'Lotus Grove' located in Colombo, Sri Lanka was used as the test network for simulations. Its topology with 63 busbars is shown in \fref{fig:network}. The number 0 node is the root node and connected to the secondary side of the MV-LV transformer. The rated capacity of the transformer is 400 kVA delta/wye and the input/output voltage rating is 11 kV/415 V. The solid lines in \fref{fig:network} represent the three-phase feeders where three-phase or single-phase loads and PV systems are connected. The transmission cable used is the Aluminium aerial bundle cable (ABC-Al/XLPE of 3x70 + N54.6 + 1x16). There are 286 single-phase or three-phase customers and 50 PV panels connected to the network. The PV panel locations are uniformly distributed across the network with assigned ratings ranging from 2-7 kW and customer peak loads assigned in the range of 0.5-1 kW through a uniformly distributed assignment process. The daily operation curves for the PV systems and the daily load profile of customer loads are shown in \fref{fig:dailyprofiles}.

\section{Results and Discussion}
\label{sec:results}
To simulate occurrences of upper and lower limit voltage violations, Monte Carlo (MC) simulations were run for three instances. \tref{tab:montecarlo} describes the three instances in terms of the hour of simulation, the network settings (PV and load setting), the number of simulations and the number of control instances that employed reactive power control or active power control (RPC or APC) or both, to mitigate voltage violations.

\begin{table}[htb]
     \centering
     \caption{Monte Carlo simulations}
     \begin{tabular}{ c c c c c c c}
     \toprule
     \multirow{2}{*}{\thead{\textbf{Time of day}}} & \multirow{2}{*}{\thead{\textbf{PV load}}} & \multirow{2}{*}{\thead{\textbf{Base load}}} & \multirow{2}{*}{\thead{\textbf{Number of}\\ \textbf{simulation runs}} }& 
     \multicolumn{3}{c}{\thead{\textbf{Control instances}}} \\
     \cline{5-7}
     & & & & RPC Q-abs & APC & RPC Q-inj \\
     \midrule
     10:00 & 76\% & 30\% & 1000 & 403 & 7 & 0 \\
     11:00 & 93\% & 50\% & 2000 & 325 & 102 & 0 \\
     21:00 & 0\% & 100\% & 500 & 0 & 0 & 500 \\
     \bottomrule
     \end{tabular}
    \label{tab:montecarlo}
\end{table}

As observed in \tref{tab:montecarlo}, voltages at simulations carried out at 10:00 violated more often, due to the low base loading. However, they tend to have a higher possibility of mitigating the upper limit violations using only RPC Q-absorption. Comparatively, at 11:00, when PV penetration increases, the rectification cannot be solely done via Q-absorption as shown in \tref{tab:montecarlo}, hence a higher number of APC rectifications (Q-absorption followed by APC) have taken place. For the simulation at 21:00 (night-time), every simulation contained lower limit violations only, which is caused by full base loading. 

With a large number of simulations using randomly generated individual base loads, PV loads, and PV positions in the network for each simulation, robustness of the control algorithm under extreme circumstances is ensured. The remaining results of the Monte Carlo simulations have been presented in \sref{subsubsec:comp_time} onwards.

\subsection{Validation of Sensitivity Matrix}
\label{subsec:sensmat results}
This section demonstrates the validity of the \sensmat\  that has been developed for the voltage sensitivity calculation for reactive power absorption/injection and active power curtailment. The validity of the \sensmat\ is analysed in terms of the accuracy and deviation of voltage sensitivity calculations for power injection/absorption and curtailment in the range 0-0.1 kW/kVar. Since the power variations within the control algorithm during optimization takes place in very small values below 0.1 kW/kVar, analysis is performed in this range. \par

\subsubsection{Deviation of Sensitivity Matrix result from load flow result}
\label{subsubsec: sensmat OPF result}
\begin{figure}[htb]
    \centering
    \includegraphics[width=0.8\textwidth, height=6cm]{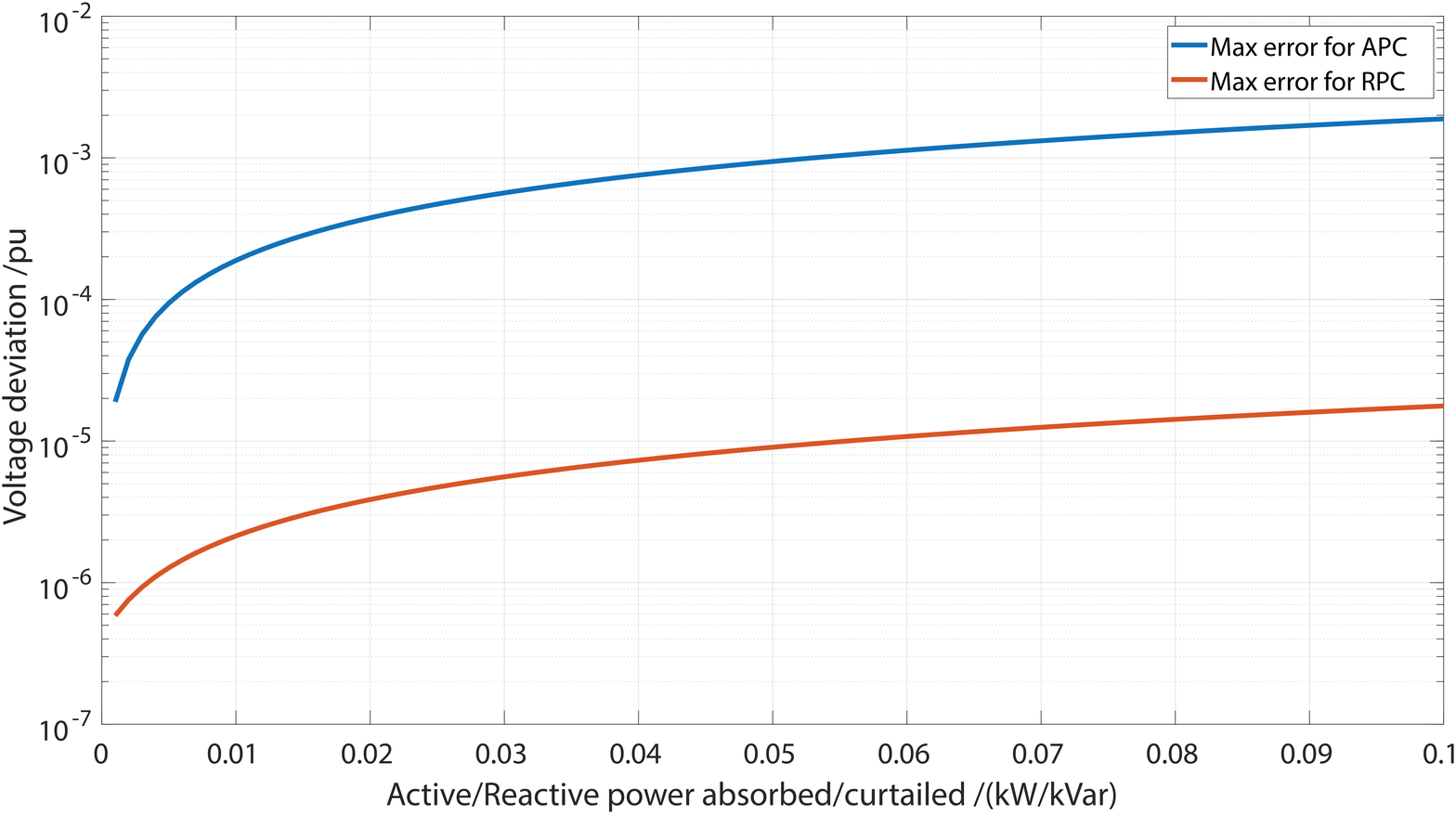}
    \caption{Deviation of voltage profile of proposed sensitivity method from load flow method}
    \label{fig:error}
\end{figure}

In this section, the maximum deviation out of all busbars in the network while using the \sensmat\ approach compared to the load flow method is analysed. It can be observed from \fref{fig:error} that maximum error for the given range of power variations is in the range of $10^{-3}$ p.u for APC and $10^-{5}$ p.u for RPC, which are negligible. 

\subsubsection{Change in voltage after APC}
\label{subsubsec: v_deviation APC}
\begin{figure}[htb]
    \centering
    \includegraphics[width=0.8\textwidth, height=6cm]{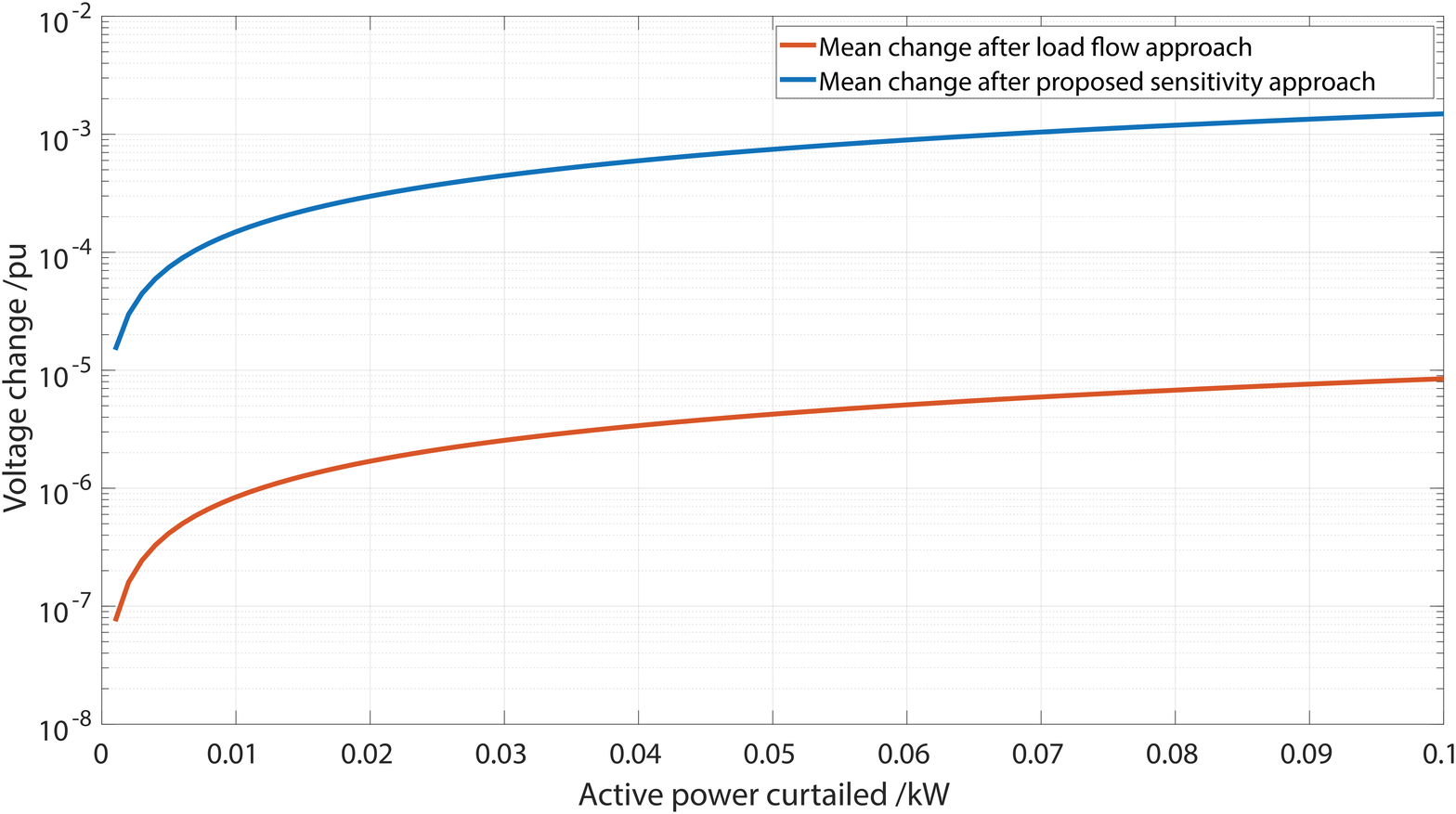}
    \caption{Change in voltage after active power control.}
    \label{fig:dV_APC}
\end{figure}

The change in voltage after active power curtailment for increasing APC amounts for the load flow approach and the \sensmat\ approach are shown in \fref{fig:dV_APC}. While there exists a deviation of the curve from the load flow results for the \sensmat\ approach, this deviation between the curves is in the range of $10^{-3}$ p.u and hence negligible. In addition, the cost function governing the optimization problem has a higher weightage for the amount of active power curtailed compared to the voltage deviation, yielding the same final voltage profile as discussed in \sref{subsec:sens_compare}.

\begin{figure}[htb]
    \centering
    \includegraphics[width=0.8\textwidth, height=6cm]{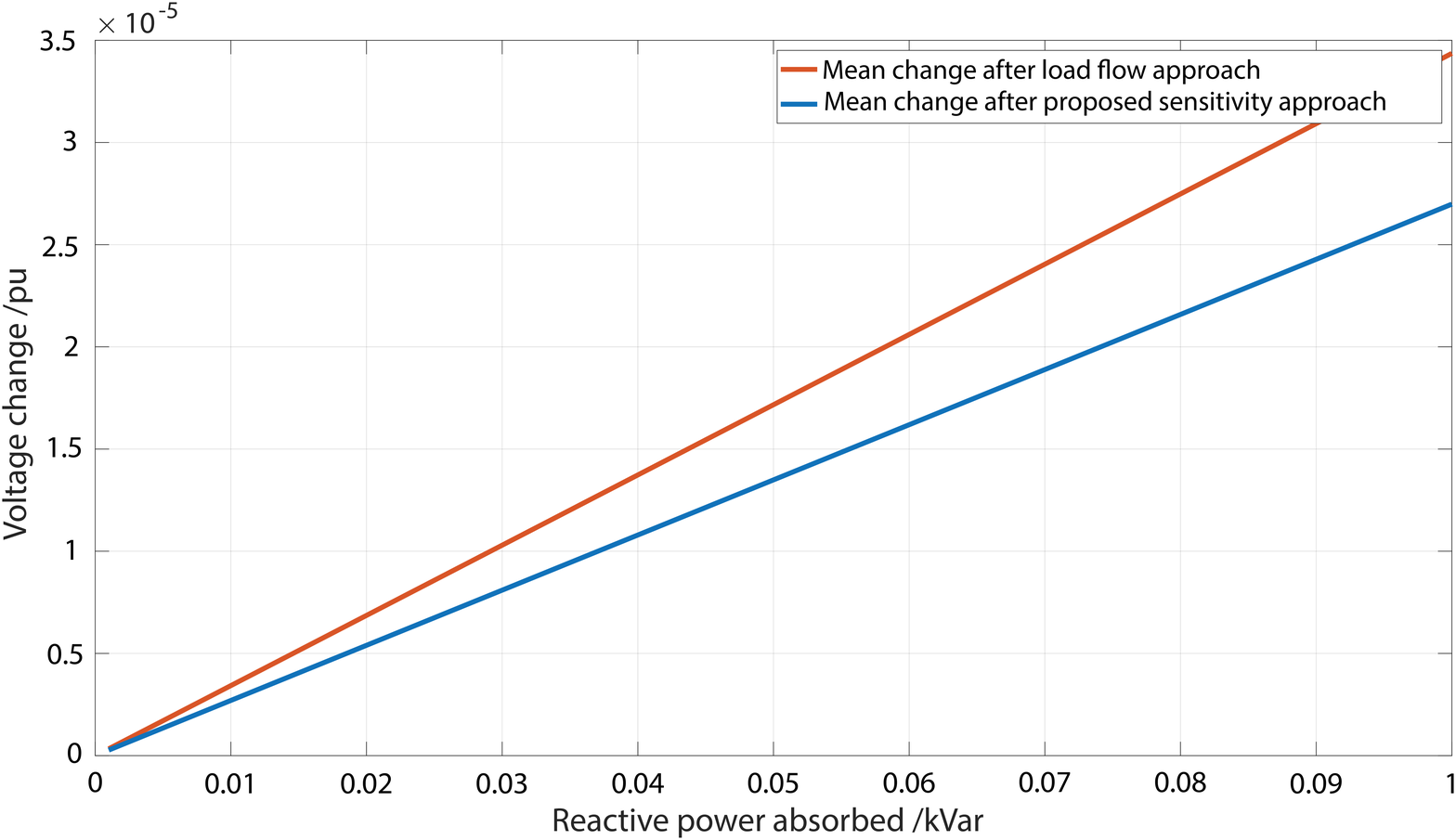}
    \caption{Change in voltage after reactive power control}
    \label{fig:dV_RPC_lin}
\end{figure}

\subsubsection{Change in voltage after RPC}
\label{subsubsec: v_deviation RPC}
The change in voltage after reactive power control for both methods is shown in \fref{fig:dV_RPC_lin}. Here it can be noted that the deviation is minimal since it is in the range of $10^{-5}$ p.u. Hence the \sensmat\ approach yields a voltage profile which is very much close to the load flow method. It is also interesting to note that the deviation between the two methods increases with the amount of reactive power injected/absorbed. Hence, if control is performed pre-emptively to mitigate large deviations from the regulatory bounds as magnitude of required RPC or APC would be lower the voltage deviation would be even more lower.

\subsubsection{Computational time comparison}
\label{subsubsec:comp_time}

\begin{figure}[H]
    \centering
    \includegraphics[width=0.8\textwidth, height=6.3cm]{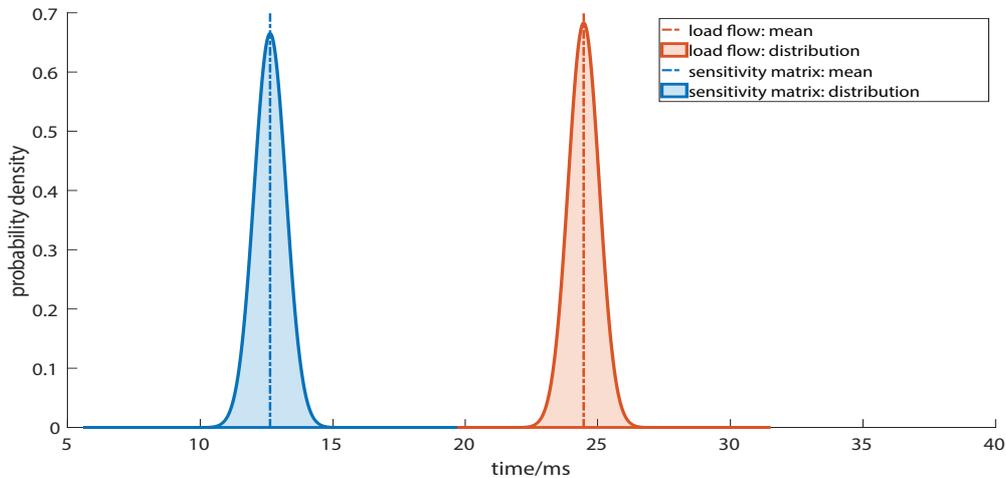}
    \caption{Comparison of computational time of a single iteration for load flow and \sensmat\ approach}
    \label{fig:single_time}
\end{figure}

\fref{fig:single_time} shows the distribution of computational time of a single iteration in both load flow and \sensmat\ approaches obtained using all 3500 MC simulations specified in \tref{tab:montecarlo}. It can be observed that the \sensmat\ approach computes the voltage almost twice as fast compared to the computational time of the load flow method. The increase in speed is validated by computing the mean computational times as in \tref{tab:comptime}, with the \sensmat\ calculation having a 48.3\% increase in speed compared to load flow.

\begin{table}[htb]
    \centering
    \caption{Elapsed computational time for sensitivity matrix and load flow calculation}
    \begin{tabular}{l  c  c  c  c}
    \toprule
    \multirow{2}{*}{\textbf{Calculation method}} & \multicolumn{4}{c}{\textbf{Computational time /ms}} \\
    \cline{2-5}
    & Mean  & Std. deviation & Minimum & Maximum\\
    \midrule
    Load Flow          & 24.5    &     0.58       &    23.7       & 35.0 \\
    Sensitivity Matrix & 12.6   &       0.60            &   12.0    & 24.1 \\
    \bottomrule
    \end{tabular}
    \label{tab:comptime}
\end{table}

\subsection{Proposed Sensitivity Matrix and Two-stage Optimization}
\label{subsec:sens_compare}

\begin{figure}[htb]
    \centering
    \includegraphics[width=\textwidth]{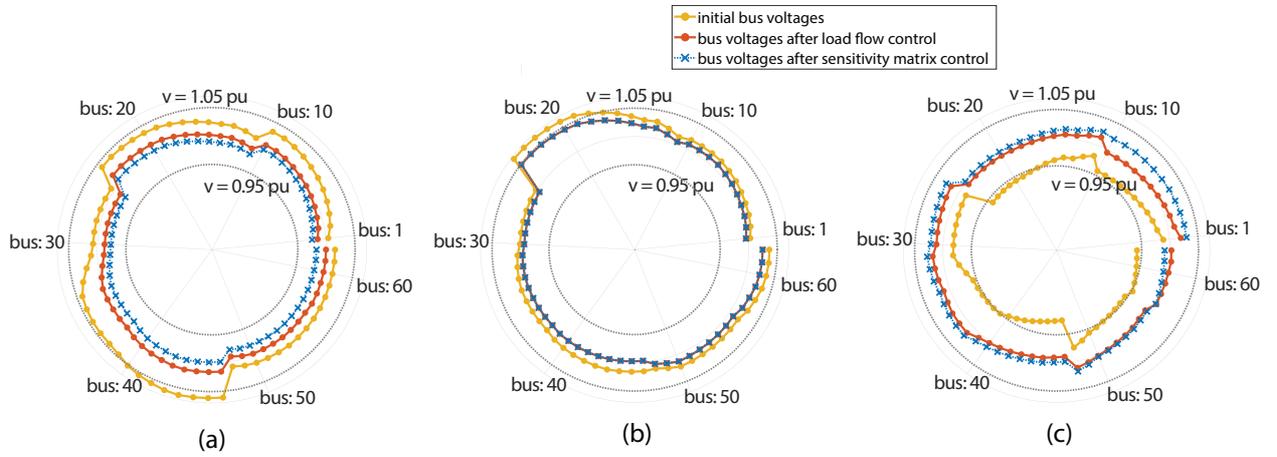}
    \caption{Voltage profiles of phase 3 before and after control at (a)10:00 using Q-absorption (b)11:00 using Q-absorption and P-curtailment, and (c)21:00 using Q-injection.}
    \label{fig:radialph3}
\end{figure}

In order to create a considerable number of violations to emulate a possible worst case scenario handling capability of the proposed \sensmat\  method and to demonstrate its robustness, PV panel positions were randomly generated such that a relatively higher number of panels were connected to phase 3, increasing the number of violations in that phase. Therefore, the initial and final voltages in the worst possible phase, phase 3, after control using \sensmat\ approach and load flow method, both using the two-stage optimization, is shown in \fref{fig:radialph3} for one particular simulation for each case at 10:00, 11:00 and 21:00. 

\begin{figure}[H]
    \centering
    \includegraphics[width=0.8\textwidth]{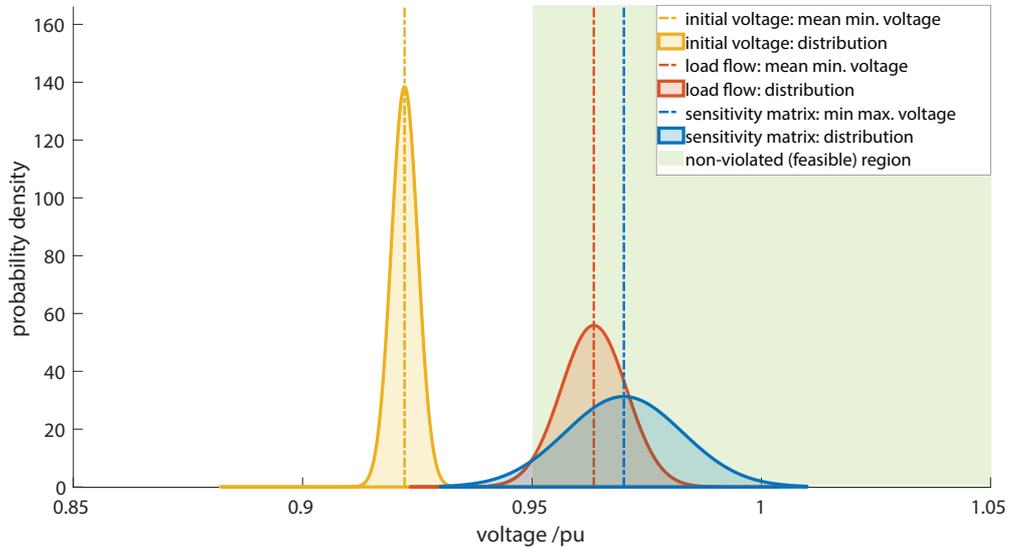}
    \caption{Distribution of minimum controlled voltages after Q injection obtained using MC simulations at 21:00}
    \label{fig:minV_QI}
\end{figure}

\begin{figure}[H]
    \centering
    \includegraphics[width=0.8\textwidth]{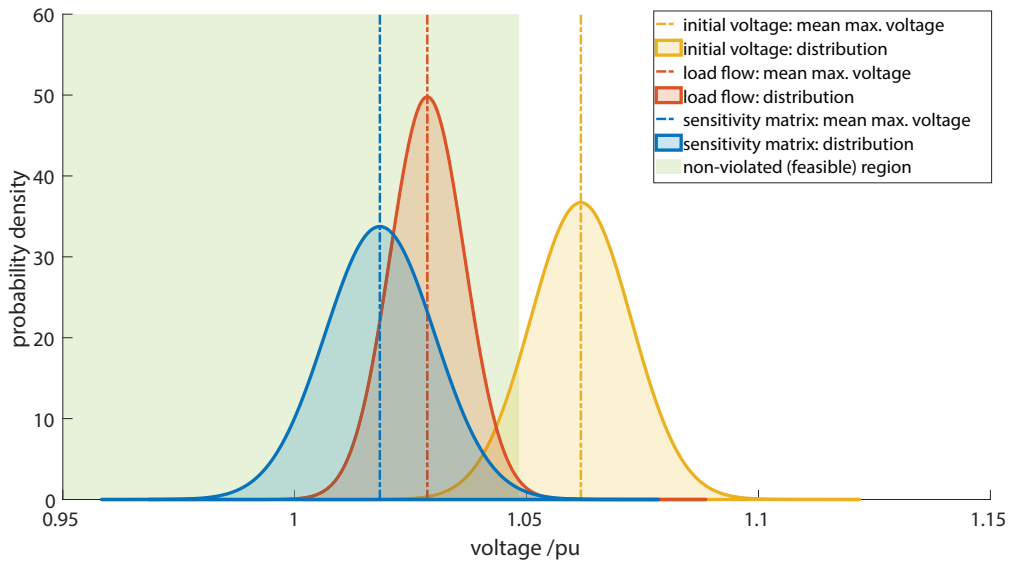}
    \caption{Distribution of maximum controlled voltages after Q absorption obtained using MC simulations at 10:00}
    \label{fig:maxV_QA}
\end{figure}

The distribution of the minimum voltage of the set of busbars after RPC using Q-injection obtained using the 500 MC simulations conducted at 21:00 as per \tref{tab:montecarlo} is shown in \fref{fig:minV_QI}. Similarly, the distribution of the maximum voltage after Q-absorption obtained using 410 MC simulations (violated simulations) out of the 1000 simulations conducted at 10:00 as per \tref{tab:montecarlo} is shown in \fref{fig:maxV_QA}. It can be observed that the lower limit violations and upper limit violations have been successfully eliminated by the control sequences. There is a slight incremental shift of $6.5 \times 10^{-3}$ p.u of the mean voltage in the minimum voltage distribution of the sensitivity matrix approach compared to the load flow approach after Q-injection (\fref{fig:minV_QI}). Similarly, the maximum voltage distribution of the sensitivity matrix method is slightly lower with a $1.02 \times 10^{-2}$ p.u  difference of the mean voltage compared to the load flow method (\fref{fig:maxV_QA}). This is due to the deviations in the sensitivity matrix approach calculated voltages for the same change in reactive power, as described in \sref{subsubsec: sensmat OPF result}. However, the optimization ensures that both methods successfully mitigate voltage violations while minimising the voltage deviations in each phase. \par

\begin{figure}[H]
    \centering
    \includegraphics[width=0.8\textwidth]{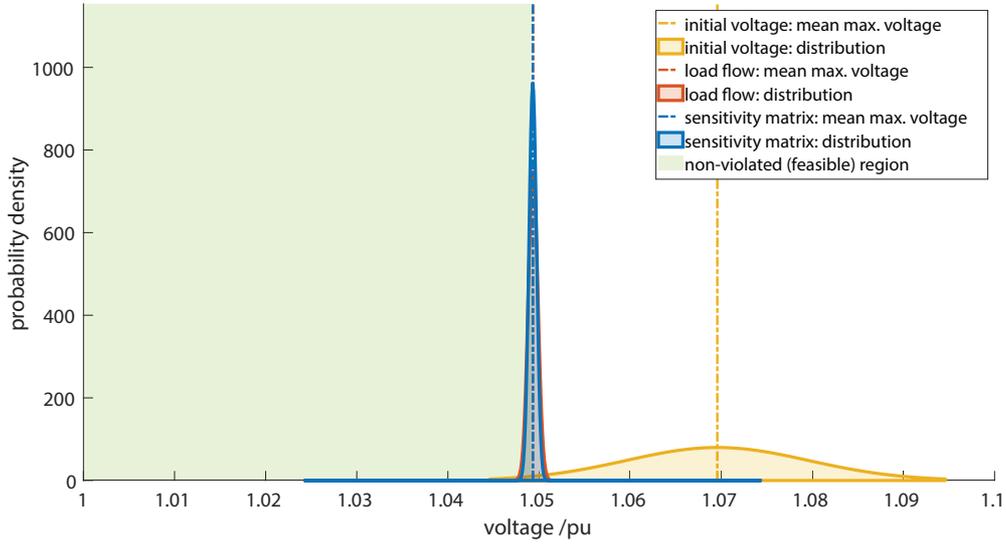}
    \caption{Distribution of maximum controlled voltages after P curtailment obtained using MC simulations at 11:00}
    \label{fig:maxV_APC}
\end{figure}

Finally, the APC algorithm is heavily dependent on the amount of active power curtailed rather than the voltage deviations in each phase. Hence, when the Q-absorption is not sufficient to remove the violations, APC is performed, which attempts to \emph{just} remove the violation. Therefore, the maximum voltage of the set of busbars is always at 1.05 p.u (the upper limit). This is seen in \fref{fig:maxV_APC} which shows the distribution of the voltage profile obtained after APC, using the 102 MC simulations (violated simulations) out of 2000 simulations conducted at 11:00 as per \tref{tab:montecarlo}. Hence in the APC case, both the sensitivity matrix approach and the load flow approach yield the exact same result. \par

\section*{Conclusion}
\label{sec: Conclusion}
In this paper, a novel Sensitivity Matrix (\sensmat) and a Centralized Active Reactive Power Management Systems (CARPMS) using inverter control to eliminate voltage limit violations are introduced. The optimum PV power settings for the control sequence are determined by a novel modified two-stage optimization algorithm. The two-stage optimization algorithm takes the predicted PV power and the estimated busbar voltages as inputs to determine the PV inverter active and reactive power settings to eliminate the voltage violations whilst minimising the unbalance in the network.\par

To demonstrate the effectiveness of the proposed \sensmat\ approach and the two-stage optimization algorithm, a simulation study is performed on an existing LV network. The PSO algorithm performance is first implemented using load flows and then using the \sensmat\ for the voltage profile generation. The results show that the \sensmat\ is able to successfully assimilate the performance of the state-of-the-art solution: the load flow; in all cases of reactive power injection, reactive power absorption and active power curtailment. The difference between the mean voltages of the proposed methodology and the load flow methods were $6.5\times10^{-3}$ p.u for RPC using Q-injection, $1.02 \times 10^{-2}$ p.u for RPC using Q-absorption. It is noteworthy that the voltage profiles obtained after APC is exactly the same for both methods (0 p.u mean voltage difference), which reaffirms the \sensmat\ approach. Furthermore, the \sensmat\ reduces the time consumed for the voltage profile generation by 48\% when compared to the load flow method. This faster inverter control will mitigate voltage violations in LVDGs thereby allowing utility providers to accommodate higher rooftop solar panels into LV networks. \par

The main advantage of the proposed two-stage optimization using the \sensmat\ is the reduction in time to generate the voltage profiles during the control sequence. Since the \sensmat\ approach is able to perform the network voltage estimation with a 48\% reduction in time with negligible accuracy loss, this will speed up the control of voltage violations in LVDGs. Furthermore, the CARPMS implementation initial cost is minimal due to the use of existing PV inverters without the need for additional device installation for the control operation. \par

Even though the proposed CARPMS focuses on the mitigation of voltage violations, it is unable to completely eliminate all violations all the time. The reactive power injection scenario is one situation, where too much reactive power injection will eliminate the lower voltage violations but give rise to upper limit violations at the other end of the network. Therefore, additional research is required to resolve this two fold issue and also to incorporate the reactive power usage into the cost function of the two-stage optimization algorithm. \par

\section*{Acknowledgements}

We would like to acknowledge the financial support provided by the National Science Foundation (NSF), Sri Lanka (research grant no. RG/2018/EA \& ICT/01) and the Peradeniya Engineering Faculty Alumni Association (PEFAA).

\clearpage

\end{document}